\documentclass[5p]{elsarticle}

\usepackage{amsmath}
\usepackage{amssymb}
\usepackage{graphicx}
\usepackage[latin1]{inputenc}
\usepackage{mathrsfs}
\usepackage[dvips]{color}

\def\sm{\,standard model}
\def\gev{\,\mbox{GeV}\,}

\newcommand{\AddrFreiburg}{
 Physikalisches Institut, Albert-Ludwigs-Universit\"at Freiburg - Fakult\"at f\"ur Mathematik und Physik, D-79104 Freiburg, Germany
 }

\begin{document}

\begin{frontmatter}
\title{A renormalization group analysis of the Hill model and its HEIDI extension}
\author[Frei]{L.~Basso}
\ead{lorenzo.basso@physik.uni-freiburg.de}

\author[Frei]{O.~Fischer}
\ead{oliver.fischer@physik.uni-freiburg.de}

\author[Frei]{J.J.~van~der~Bij}
\ead{vdbij@physik.uni-freiburg.de}

\address[Frei]{\AddrFreiburg}

\begin{abstract}
The parameter space of the simplest extension of the standard model is studied in the light of the $125$\gev Higgs boson discovery. The Hill model extends the scalar sector of the standard model with a real singlet, that mixes with the {\sm} Higgs boson. The two-loop {\sm} renormalization group equations are completed with the one-loop Hill equations. Stability up to the Planck scale is possible without tension with the other parameters. An extension with more singlet fields, in particular a higher-dimensional (HEIDI) field, is presented.
\end{abstract}

\end{frontmatter}

\section{Introduction}
\label{sect:intro}

The standard model describes particle physics in great detail. Nonetheless in the past most work has been based on the assumption, with varying motivations, that the standard model must be incomplete and that new physics should be just around the corner. With the new data from the LHC it seems reasonable to question this assumption. The fact that the LHC has found no evidence for new physics puts strong constraints on possible extensions. For instance, the fact that LHCb finds full agreement with standard model predictions basically rules out any new form of flavor physics at accessible scales. Also direct searches have found no sign for new physics,  deep into the TeV scale. 

The analysis of the Higgs boson properties is an eminent goal of the physics at the LHC. The data show a resonance at 125\gev~that is consistent with a standard model Higgs boson~\cite{Higgs:exp}. However there is still a considerable uncertainty on the value 
of the overall coupling strength to the different particles in the theory. 
Therefore constraints
on the presence of extra states mixed with the Higgs boson are relatively mild.

The two-loop renormalization group equations (RGEs) of the {\sm} with the present central value for the Higgs mass do not lead to a theory 
that is stable up to the Planck scale~\cite{Bezrukov:2012sa,Degrassi:2012ry,Branchina:2013jra} (see, however,~\cite{Alekhin:2012py,Masina:2012tz}). This statement is strongly dependent on the value of the top quark and Higgs boson masses. Based on old ideas, a number of studies in the light of recent developments have reinforced 
the picture that simple scalar extensions can relax this tension and lead to a theory that is stable up to very large mass-scales, either via tree-level mixing~\cite{Extra_scalars,Recent_singlets,Basso:2013vla} or as new thresholds in the running of the quartic coupling~\cite{Lebedev:2012zw}.

Another major point of interest in the running of the Higgs coupling is the fact,
that the vanishing of the Higgs quartic coupling at the Planck scale 
sets the lower bound on the Higgs mass when related to inflation
 (see, e.g.,~\cite{Bezrukov:2012sa,vanishing_lambda} and references therein).

In the present Letter we investigate at the two-loop level whether the Hill model can improve the validity range of the \sm. Further, we begin the analysis of its extension, the HEIDI model, by constraining the parameter space of the latter compatible with inflation in view of the Higgs boson discovery at the LHC.

This Letter is organized as follows: in the next section we describe the Hill model and give the one-loop RGEs of the model parameters. In section~\ref{sect:results} we present the results of our analysis. In section 4 we discuss the HEIDI extension and in the last section, we conclude.

\section{The model}
\label{sect:model}

The Hill model is the simplest extension of the scalar sector of the \sm. One real scalar field ($H$) is introduced, that mixes with the {\sm} field ($\Phi$)~\cite{Hill:1987ea, Basso:2012nh}. The scalar potential reads
\begin{equation}
{\cal V} =  \lambda_1 \left( \Phi^\dagger \Phi - \frac{v^2}{2}\right) ^2
 +\lambda_2 \left( \sqrt{2} f_2 H - \Phi^\dagger \Phi \right) ^2 \, .
\label{eq:Vtree}
\end{equation}

A defining feature of the Hill model is the absence of $H$ self-interactions, as well as $H^2\Phi^2$ terms, 
so there are only two new parameters. The fact that the extra interaction is superrenormalizable leads to
a limited impact  on the renormalization group, whose study is the aim of this Letter. The beta function for the \sm~Higgs coupling stays untouched, which makes it possible to study it at $2$ loops while still considering new physics at $1$ loop, as discussed in the next section.
Moreover, the absence of $H$ self-interactions is a key feature that allows for an extension to higher-dimensional (HEIDI) fields in a renormalizable way up to six dimensions, to be discussed in section~\ref{Sec:HEIDI}.
Notice that the normalization of the couplings $\lambda_{1,2}$ is changed by a factor of $2$, compared to Ref.~\cite{Basso:2012nh}, in order to conform with the conventions in~\cite{Degrassi:2012ry}.

In unitary gauge, $\Phi=\left(0, (h + v)/\sqrt{2}\right)^T$, and $H= (h' + v_H)/\sqrt{2}$. The scalar potential is minimized for $\left< \Phi \right> = v = 246$ \gev and $\displaystyle \left< H \right> = \frac{v^2}{2 f_2} \equiv v_H$. At the minimum, the
two {\it CP}-even scalars mix as follows:
\begin{equation}
\left( \begin{array}{c} h_1\\ h_2 \end{array} \right) = \left( 
\begin{array}{cc} c_{\alpha} & s_{\alpha}\\ - s_{\alpha} & c_{\alpha} \end{array} \right) \, \left( \begin{array}{c} h \\ h' \end{array} \right)\,,
\end{equation}
with $s_\alpha\,(c_\alpha)$  the sine (cosine) of the mixing angle $\alpha$.
The mass eigenstates $h_{1(2)}$ couple to the {\sm} particles with an overall $c_{\alpha}(s_{\alpha})$ prefactor and have masses
\begin{equation}\label{h_mas}
m_{h_{1,2}}^2 = \left( \lambda_2 f_2^2 + \lambda_3 v^2\right) \pm
\sqrt{4 \lambda_2^2 v^2 f_2^2 + (\lambda_2 f_2^2 - \lambda_3 v^2)^2}\, ,
\end{equation}
where we have defined $\lambda_3 = \lambda_1 + \lambda_2$ and $m_{h_1} < m_{h_2}$. Notice that $\lambda_3$ is the total self coupling of the Higgs doublet.
The mixing angle is given by
\begin{equation}\label{h_ang}
c_{\alpha}^2 = \frac{m^2_{h_2}-2 \, \lambda _3 v^2}{m^2_{h_2} - m^2_{h_1}}\, .
\end{equation}

Eqs.~(\ref{h_mas}) and~(\ref{h_ang}) can be inverted to express the parameters $f_2$, $\lambda_2$ and $\lambda_3$ in terms of the observable scalar masses and mixing angle,
\begin{eqnarray}
\lambda_2 &=&  \frac{s^2_\alpha c^2_\alpha (m^2_{h_2} - m^2_{h_1})^2}{2\,v^2
(s^2_\alpha m^2_{h_1}+c^2_\alpha m^2_{h_2} )}\, , \label{eq:lambda2}\\
\lambda_3 &=&  \frac{c^2_\alpha m^2_{h_1} + s^2_\alpha m^2_{h_2}}{2\,v^2} \,, \label{eq:lambda3}\\
f_2 &=& v \frac{s^2_\alpha m^2_{h_1} + c^2_\alpha m^2_{h_2}}{s_\alpha c_\alpha (m^2_{h_2} - m^2_{h_1})}\, .\label{eq:f2}
\end{eqnarray}

To extract the equations describing the running of the parameters, it is easiest to consider the one-loop improved scalar potential. The presence of the Hill singlet modifies the contribution solely of the real Higgs field, via tree-level mixing, while leaving all the rest untouched. It is therefore sufficient to replace the real Higgs contribution in the one-loop scalar potential with the following term:
\begin{equation}
{\cal V}_{\rm Hill}^{(1)} = \frac{1}{64 \pi^2}{\rm Tr}\left[\left(M_\varphi^2\right)^2\left[ \ln\left(\frac{M_\varphi^2}{\mu^2}\right) - {3 \over 2} \right]\right] \,,
\end{equation}
in which the field-dependent mass now reads as a matrix;
\begin{equation}
M_\varphi^2 = \left( \begin{array}{cc} \left(-m_0^2 + 3 \lambda_3 \varphi^2 - \eta H\right) & -\eta \varphi \\ -\eta \varphi & m_1^2 \end{array} \right)\,,
\end{equation}
with $\varphi^2=\Phi^\dagger\Phi$ being a real field. The condition of minimization of the tree-level potential fixes the parameter 
$m_0^2 =  \lambda_1 v^2$, and we defined
\begin{eqnarray}
m_1^2 & = & 2 \, \lambda_2 \, f_2^2 \,,\\
\eta & = & 2 \, \lambda_2 \, f_2 \,.
\end{eqnarray}
Imposing the condition that the improved potential does not depend on the renormalization scale yields the RGEs for $m_1,\, \eta$
\begin{eqnarray}
\frac{1}{m_1^2}\frac{ {\rm d}m_1^2}{{\rm d} t} & = & \frac{\eta^2}{16 \pi^2 m_1^2} \,,\\
\frac{ {\rm d}\eta}{{\rm d} t} & = & -2 \gamma_0 \eta + \frac{3 \eta \lambda_3}{16 \pi^2}\, ,
\end{eqnarray}
with $t=\ln{Q}$ and $\gamma_0 = -\frac{1}{16 \pi^2}\left( 3 y_t^2 - {9 \over 4}g^2 - {3\over 4}g'^2\right)$ the anomalous dimension of the Higgs boson at one loop.
 $g,g'$ are the gauge couplings and $y_t$ is the top Yukawa coupling. Among the RGEs of the {\sm} at one loop, only the equation for the Higgs mass receives a contribution from the Hill field,
\begin{equation}\label{eq:m0}
\frac{1}{m_0^{2}} \frac{ {\rm d}m_0^2}{{\rm d} t} = -2 \gamma_0 + \frac{2}{16 \pi^2} \frac{\eta^2}{m_1^2}\,.
\end{equation}
The other equations for gauge and Yukawa couplings remain unchanged. Finally and most importantly, the absence of a quartic coupling between the Hill field and the {\sm} doublet ensures that the RGE  for $\lambda_3$ reads as in the \sm.

\section{Results}
\label{sect:results}

We identify $h_1$, the lighter mass eigenstate, with the scalar boson as observed at the LHC, with a mass of 125 \gev. The mass of the second boson and the mixing angle $\alpha$ are our free parameters. We have thus
\begin{eqnarray}\label{h1-h2}
m_{h_1} = 125 \text{\gev} & \mbox{ with } & m_{h_2} > 125 \text{ \gev}\,.
\end{eqnarray}
We will call the particle with mass of 125 \gev the \sm-like Higgs boson, and the other one the Hill field, with mass $m_{\rm Hill}=m_{h_2}$. The labels $h_1$, $h_2$ and their corresponding masses will be employed accordingly. 

\begin{figure}
\begin{center}
\includegraphics[width=0.35\textwidth,angle=-90]{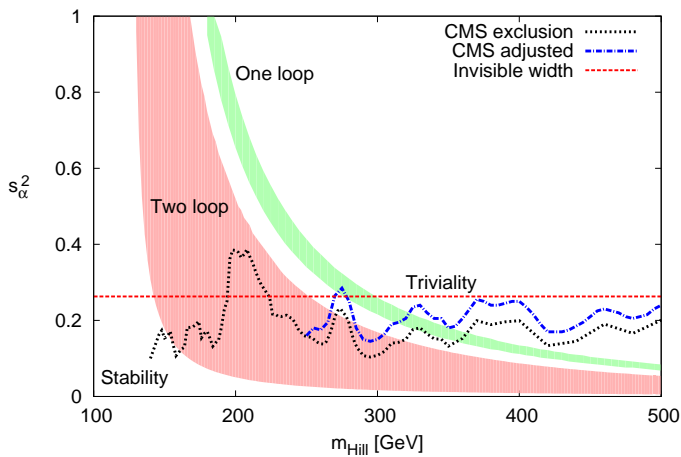}
\end{center}
\caption{Allowed parameter space in the Hill model. The green(red) area corresponds to parameter sets yielding a stable and non-trivial potential up to the Planck scale, considering one-(two-)loop RGEs for the {\sm} parameters. The blue/dashed line represents the experimental exclusion, adjusted for the Hill model (see text for details).}
\label{fig_Hill-2}
\end{figure}

The RGEs from the previous section are solved numerically, identifying the regions in parameter space that comply with the following conditions:
\begin{equation}\label{cond_1}
0 < \lambda _{3}(Q) < 1 \qquad \forall \; Q \leq Q'\, ,
\end{equation}
where the left-/right-hand side is usually referred to as the ``vacuum stability''/``triviality'' condition. Their meaning is that the 
condition of perturbativity and the existence of a well-defined  vacuum of the model must be fulfilled at any scale $Q\leq Q'$, with $Q'$ the ultimate scale of validity of the theory. In this Letter, we take $Q'=M_{Pl}=10^{19}$ \gev. Both conditions have been chosen  somewhat restrictive in order to show that the model under consideration is a simple {\sm} extension that is viable up to the Planck scale,
and to identify the tightest possible bounds in parameter space. Relaxing these conditions (e.g., considering $Q'<M_{Pl}$ or increasing the perturbativity bound) will lead to results for which our findings are a valid subset.
The allowed parameter space is presented in Fig.~\ref{fig_Hill-2}. 

As remarked earlier, the RGE for the quartic coupling reads as in the {\sm}. However, the {\sm}~Higgs boson is now mixed with the Hill field, which will affect the former running. Strictly speaking, a threshold effect appears at the mass of the new particle (see for instance Ref.~\cite{Lebedev:2012zw}), which should be treated with care. If the difference between the mass threshold and the initial energy scale is much smaller than the
range in energy that is considered, one could in first approximation use for all energies the modified RGEs, since the numerical difference with and without threshold effect is very small. In this Letter we consider Hill boson masses below 1 TeV, and evolve the RGEs all the way up to the Planck scale: it is then a justified approximation to consider the impact of the Hill boson on the boundary condition for the {\sm}~quartic coupling instead of the full treatment of the threshold effect.
Hence, in first approximation, it is possible to improve on the previous study by considering the two-loop RGEs of the {\sm} integrated by the one-loop equations for the Hill parameters from the previous section, and consider the effect of the Hill boson on the {\sm} solely in Eq.~(\ref{eq:m0}) and in the modified boundary condition for the {\sm} quartic coupling. At the electroweak scale the following two-loop corrected Higgs coupling is used~\cite{Degrassi:2012ry}:
\begin{equation}\label{delta_lambda}
\lambda_3 = 0.12577 + v_H \,\delta_H + v_t\,\delta_t +\Delta\lambda\,,
\end{equation}
with the Hill-induced tree-level shift ($\Delta \lambda$) from Eq.~(\ref{eq:lambda3}) and the other parameters given by, respectively,
\begin{eqnarray}\label{shift}
\Delta \lambda & = & s_\alpha^2 \frac{m_{h_2}^2 - m_{h_1}^2}{2\,v^2}\,, \\
v_H = 0.00205\,, && \delta_H = \frac{m_h}{\rm \gev}- 125.0 \,, \\
v_t = 0.00004\,, && \delta_t = \frac{m_t}{\rm \gev}- 173.15 \,, 
\end{eqnarray}
where the indices $H$ and $t$ denote the Higgs and top parameters. Furthermore, we adopt $\sigma_t=|m_t- 173.15|$ \gev $=0.7$ \gev and $\sigma_H=|m_h- 125.0|$ \gev $=1.0$ \gev as one-sigma standard deviations. The shift on the boundary condition of Eq.~(\ref{shift}) is completely general, see~\cite{Basso:2013vla}, and therefore our results can be considered as  general  for models in which the quartic coupling running is not affected by the extra content of the model, but only its boundary condition.

Fig.~\ref{fig_Hill-2} shows the range of parameters allowed by the stability and triviality conditions for $m_t = 173.15$ \gev and $m_h= 125.0$ \gev.
Comparing the two-loop case to the one-loop one, it is evident that a smaller tree-level shift is sufficient to achieve stability up to the Planck scale. 

We compare the previous results to the experimental constraints, given by CMS~\cite{Chatrchyan:2013yoa}. We point out that these bounds come from the standard Higgs searches applied to the heavy scalar. This means that the heavier scalar has all the properties of the {\sm}~boson with a higher mass. However, no new decay channels were
considered, such as the decay of the heavy state into pairs of the lighter one.
The blue/dashed line in Fig.~\ref{fig_Hill-2} is the exclusion curve, adjusted
to account for the new $h_2 \to h_1h_1$ channel, kinematically accessible for $m_{h_2} \gtrsim 250$ \gev, as follows.

Experimental data are usually presented in terms of excluded ratio of cross sections at a given mass: $\kappa(m_{h_2}) \equiv \sigma^{DATA}(m_{h_2}) / \sigma^{SM}(m_{h_2})$. In the Hill model,
\begin{equation}\label{kappa_factor}
\kappa (m_{h_2},\,s_\alpha) = \frac{s_\alpha^4 \Gamma_{\rm SM}^{\rm tot}(m_{h_2})}{\Gamma_{\rm Hill}^{\rm tot}(m_{h_2})}\,,
\end{equation}
given that each partial width into {\sm} particles is reduced by $s^2_\alpha$.
$\Gamma^{\rm tot}_{\rm SM}(m_{h_2})$ is the total decay width for a {\sm} Higgs boson of mass $m_{h_2}$. 
The total decay width of the Hill particle is given by:
\begin{equation}
\Gamma_{\rm Hill}^{\rm tot}(m_{h_2}) = s_\alpha^2 \Gamma^{\rm tot}_{\rm SM}(m_{h_2}) + \frac{F^2(\alpha,m_{h_2})}{32 \pi m_{h_2}}\sqrt{1-\frac{4 m_{h_1}^2}{m_{h_2}^2}}\, .
\end{equation}
where $F$ is the $h_1-h_1-h_2$ coupling,
\begin{equation}
F(\alpha,m_{h_2})=s_\alpha c^2_\alpha (2 m_{h_1}^2 + m^2_{h_2})/v\,.
\end{equation}
When $BR(h_2\to h_1h_1) = 0$ (i.e., $m_{h_2}<2\,m_{h_1}$ or $s_\alpha = 0$), $\kappa = s^2_\alpha$, as usually assumed in the experimental analyses.

Solving Eq.~(\ref{kappa_factor}) for $s^2_\alpha$ gives the adjusted exclusion curve in the $(s^2_\alpha$--$m_{\rm Hill})$ plane of Fig.~\ref{fig_Hill-2} (the blue/dashed line), compared to the naive interpretation of the exclusion limit when $h_2$ decays only into {\sm} particles (black/dotted line). It is clear that such naive interpretation is not adequate.

The observation of the $125$\gev Higgs boson also indirectly constraints the scalar mixing angle in this model. 
 By simply averaging the overall Higgs strengths of ATLAS ($\mu = 1.30\pm 0.2$) and CMS ($\mu = 0.80 \pm 0.14$)~\cite{Higgs:exp}, we obtain a $2\sigma$ constraint on the 
Higgs modulation of $\sin^2{\alpha} \lesssim 0.263$, compatible with constraints from fitting an invisible width~\cite{invisible}. We show the bound in Fig.~\ref{fig_Hill-2}.  It is clear that for Hill boson masses above $250$ \gev, the direct search limit sets the most stringent bound.

\begin{figure}
\begin{center}
\includegraphics[width=0.37\textwidth]{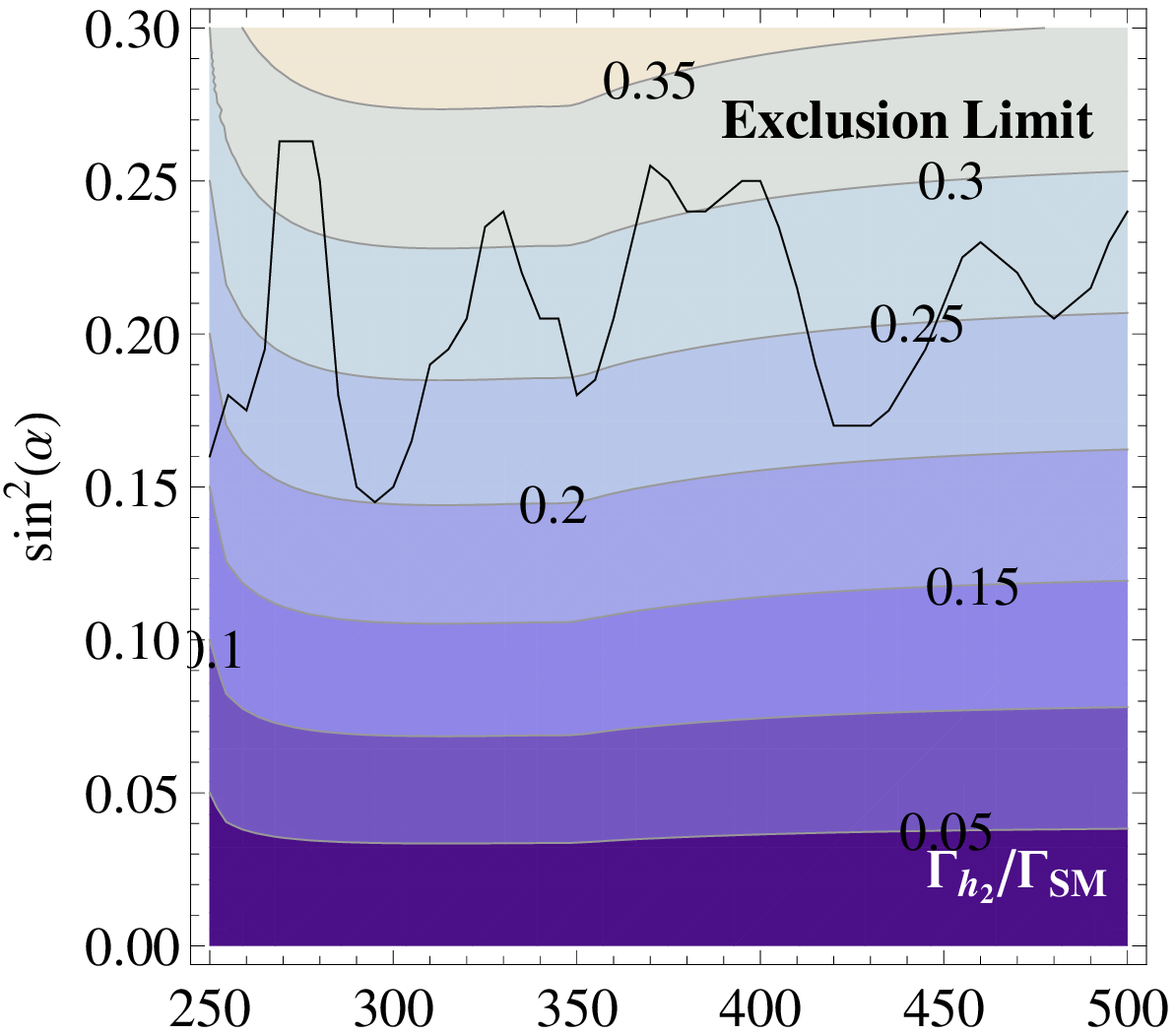}\\
\vspace{0.1cm}
\includegraphics[width=0.37\textwidth]{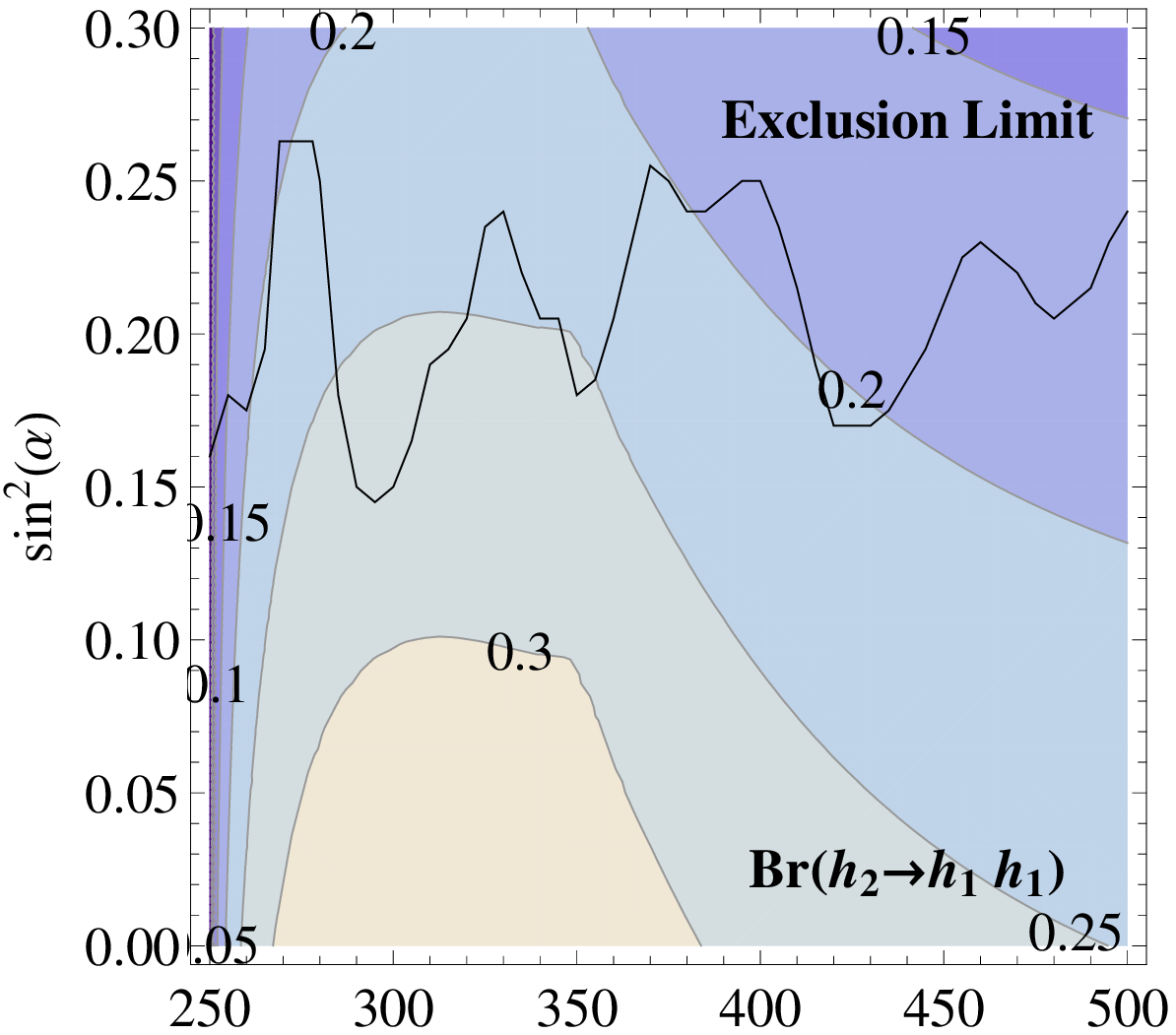}\\
\vspace{0.1cm}
\includegraphics[width=0.37\textwidth]{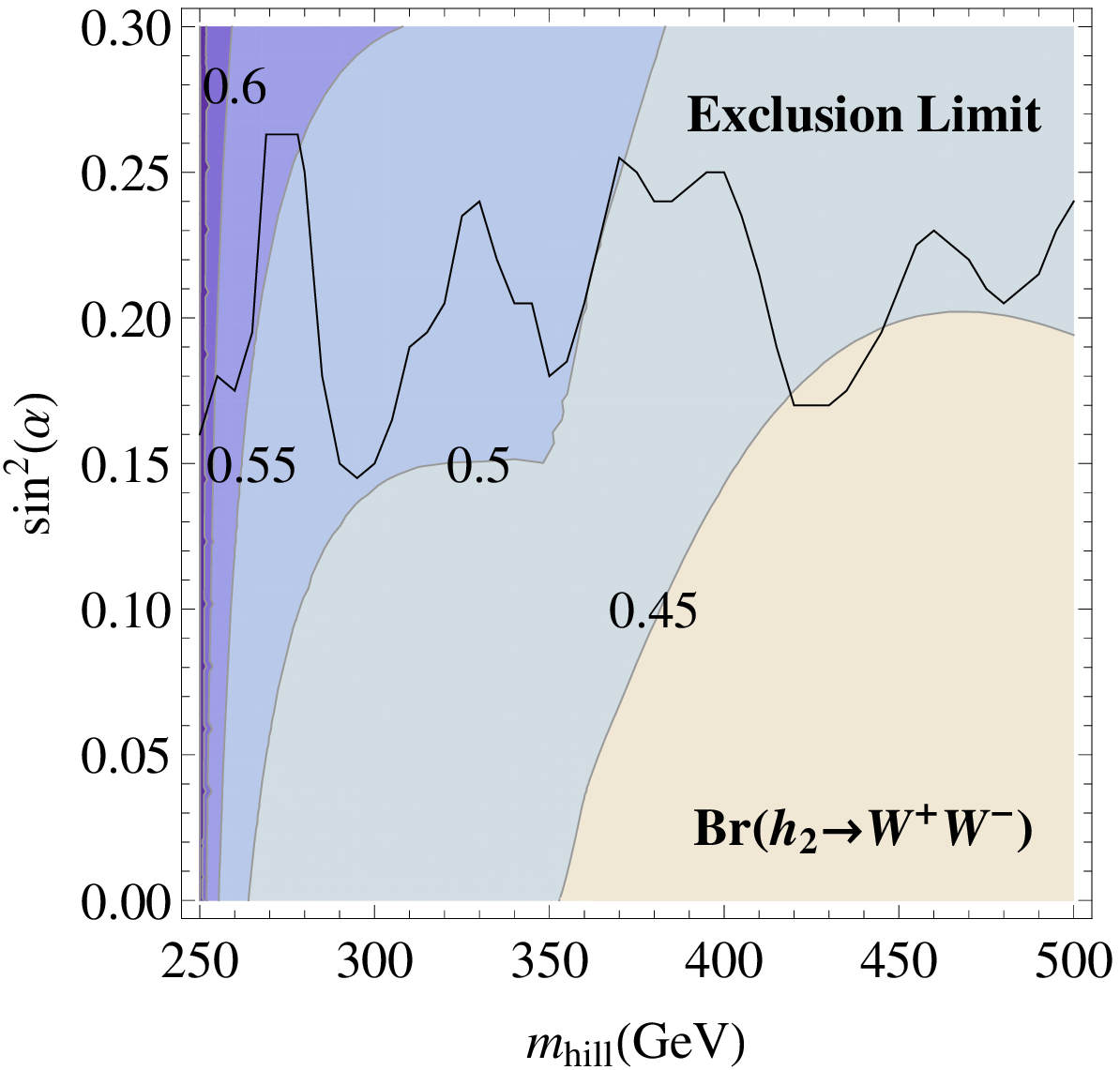}
\end{center}
\caption{Top panel: ratio $\Gamma_{\rm Hill}^{\rm tot}/\Gamma^{\rm tot}_{\rm SM}$. Middle panel: $BR(h_2 \to h_1h_1$). Bottom panel: $BR(h_2 \to W^+ W^-)$. The solid line is the adjusted experimental exclusion.}
\label{fig_width}
\end{figure}

The total Hill boson decay width is shown in the top panel of Fig.~\ref{fig_width}. The partial widths were calculated with \texttt{HDECAY}~\cite{Djouadi:1997yw}.
Within the Hill model the total decay width is always less than the standard model width.
The other panels of the same figure show the branching ratios (BR) of the Hill boson into pairs of the $125$ \gev Higgs bosons and into $W^+W^-$ bosons pairs, respectively. The BR into $ZZ$ pairs
is given by the {\sm} ratio to  the $WW$ case and therefore not shown. 
Furthermore, for small angles, its total decay width can be much smaller than the {\sm} case for the same masses, which means that the experimental analyses (often assuming the {\sm} Higgs width) might account for more background than required, as the real peak
is  much narrower than assumed.

\begin{figure}
\begin{center}
\includegraphics[width=0.35\textwidth,angle=-90]{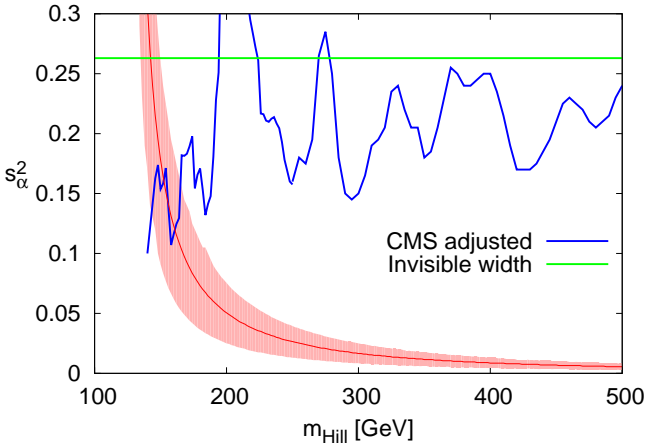}
\caption{Regions in the ($s_\alpha^2$-$m_{\rm Hill}$) plane such that the scalar quartic coupling vanishes at the Planck scale, varying the top quark and Higgs boson masses by one sigma around their central value (represented by the red/solid curve inside the shaded areas).}
\label{fig_dMhMt}
\end{center}
\end{figure}

The request of a vanishing Higgs quartic coupling at the Planck scale is of particular interest. In Fig.~\ref{fig_Hill-2}, this request represents the lower bound on the allowed parameter space, here only for the Higgs boson and top quark central value of the respective masses. It is therefore interesting to evaluate the quantitative effect of a one-sigma variation of these two parameters. This is done in Fig.~\ref{fig_dMhMt}, where the solid line represents the central value. 

For completeness, the required range of $\Delta \lambda$ from Eq.~(\ref{delta_lambda}), depicted in Fig.~\ref{fig_Hill-2}, is given
in a model-independent fashion by:
\begin{equation}
0.005 \leq \Delta \lambda \leq 0.105\, ,
\end{equation}
where the lower bound comes assuming $m_t= 172.45$ \gev, $m_{h_1}=126.0$ \gev and the upper bound assuming $m_t=173.85$\gev, $m_{h_1}=124.0$\gev. 

The variation for the triviality bound (represented by the upper edge of the colored areas in Fig.~\ref{fig_Hill-2}) is not shown, since it varies below the percent level, when the Higgs boson and the top quark masses are varied within one sigma around their respective central values.

The vanishing of the scalar quartic coupling at a very large scale $\mu$ is not the only interesting aspect one might consider. A more restrictive condition on the Higgs mass comes from the request that also its beta function vanishes at some scale $\mu$,
\begin{equation}\label{cond_vanish}
\lambda(\mu) = \beta_\lambda(\mu) = 0\, .
\end{equation}
This situation is of interest because it plays an important role in the argument
whether the Higgs boson might be responsible for the inflationary phase of the universe, where said equation sets the minimum Higgs mass  compatible with slow-roll inflation. 
We focus on Eq.~(\ref{cond_vanish}) also because it might be related to  the existence of a conformal window for scales near the Planck mass.
The solution of Eq.~(\ref{cond_vanish}) is found in Ref.~\cite{Bezrukov:2012sa} to be
\begin{eqnarray}\label{mass_vanish}
\hskip -0.3cm
M_{min} &=& \left[ 129.57 + 2.2\, \delta_t-0.56\, \delta_{\alpha_s} \pm \delta_{th}\right]
\, \scriptscriptstyle \mbox{\gev} ,\\
\delta_t &=& \frac{(m_t/\mbox{\gev}-172.9)}{1.1}\, ,\\
\delta_{\alpha_s} &=& \frac{\alpha_s-0.1184}{0.0007}\, ,\\
\delta_{th} &\simeq& 1.2\, .
\end{eqnarray}
where $M_{min}$ is the {\sm} Higgs mass where at some scale $\mu$ (close to the reduced Planck mass, $2.44~\cdot~10^{18}$ \gev), both the scalar coupling and its derivative vanish. We can interpret the equation in the Hill model by considering $M^2_{min}\equiv 2\lambda_3 v^2$ in Eq.~(\ref{eq:lambda3}), to be solved for the Hill mass or, equivalently, for the mixing angle. In a model independent fashion,  the required $\Delta \lambda$ from Eq.~(\ref{delta_lambda}), is $\Delta \lambda \simeq 0.0096$ for the central values of the top quark and Higgs boson masses.

\section{Extension to the HEIDI model}
\label{Sec:HEIDI}

The Hill model can be easily extended to include more Hill fields $H_i$.
In terms of the modes $H_i$ the Lagrangian, that we use, is the following:
\begin{eqnarray}
\label{lagrangian2}
\mathcal{L} &=&-D_\mu \Phi^\dagger D^\mu \Phi -\frac{1}{2} \sum (\partial_\mu H_k)^2 \nonumber \\
&& +m_0^2 \Phi^\dagger \Phi -\lambda(\Phi^\dagger \Phi)^2-\frac{1}{2}\sum m_k^2 H_k^2\nonumber \\
&& - g  \Phi^\dagger \Phi \sum H_k-\frac{\zeta}{2} \sum H_i H_j. ~~~
\end{eqnarray}

We take the coupling of the fields $H_i$  to the Higgs field to be equal.
This condition is respected by renormalization because of the permutation symmetry for the
fields $H_i$, that is only softly broken by the mass terms. This property allows  us to take
the limit of an infinite number of fields in the form of a higher-dimensional (HEIDI) Hill field~\cite{vanderBij:2006ne,vanderBij:2011wy}.
To go to a higher-dimensional field one simply takes
$m_k^2=m^2+ m_{\gamma}^2\,\vec k^2$, where $\vec k$ is a $\gamma$-dimensional
vector, $m_{\gamma}=2\pi/L$ and $m$ is a $d$-dimensional mass term for the field $H$.
Subsequently one can take the continuum limit, which corresponds to the Hill field moving in a flat
open space.  Such a theory is renormalizable, as long as the dimension of the
space is six or less. The Higgs propagator is given by:

\begin{eqnarray}
\label{higgsprop4}
D_{HH}(q^2)&=&\Bigg(q^2 +M^2-\frac{g^2 v^2 \sum (q^2+m^2_i)^{-1}}{1 + \zeta \sum (q^2+m^2_i)^{-1}} \Bigg)^{-1}
\nonumber \\
&=& \sum \frac{c_i^2}{q^2+M_i^2}\, ,
\end{eqnarray}
where $M^2=2 \lambda v^2$.

In the continuum limit the propagator can be rewritten in the form \cite{vanderBij:2011wy}
\begin{eqnarray}
\label{higgsprop5}
D_{HH}(q^2)=\Bigg(q^2 +M^2-\frac{\mu^{8-d}}{(q^2+m^2)^{\frac{6-d}{2}} \pm \nu^{6-d}} \Bigg)^{-1}
\end{eqnarray}
where $\nu$ is positive and the sign in front
of the $\nu$ term is the sign of $\zeta$.

We first want to prove that the coefficients $c_i^2$ sum up to one. Performing a contour integral on Eq.~(\ref{higgsprop4}), with the contour at infinity, we get the residues of the poles of both sides,
\begin{equation}\label{proof_aux}
1 = \sum_i c_i^2\,.
\end{equation}
In general this is a consequence of diagonalizing the scalar mass matrix with a unitary matrix. For the Hill model, with one extra scalar field, the $c_i$ are simply the sine and cosine of the scalar mixing angle.

Subsequently we want to generalize Eq.~(\ref{eq:lambda3}). 
Performing once more a contour integral at infinity for the following function:
\begin{equation}
q^2 D_{HH}(q^2) - 1 =   \sum_i \frac{q^2 c_i^2}{q^2+M_i^2} - 1\,,
\end{equation}
one finds
\begin{equation}\label{general_lambda}
M^2 = \sum_i c_i^2 M^2_i\,.
\end{equation}

In the continuous case, the RHS of the above equation must be replaced with an integral over the spectral density,
where poles are still allowed as delta functions in the spectral density.
 This  leads to the following result for the Higgs coupling $\lambda$:
\begin{equation}\label{lambda_gen}
\lambda = \frac{\int_0^\infty s\,\rho(s) ds }{2 v^2}\,.
\end{equation}\\

Eq.~(\ref{lambda_gen}) allows for the reinterpretation of  the results for the Hill model in its HEIDI extension. This reinterpretation is however not completely straightforward because of the additional parameters in the model.
In fact, within the HEIDI models various possibilities regarding the spectrum of the Higgs propagator exist.
Dependent on the parameters there are zero, one or two peaks plus a continuum, that would show up as an invisible decay spectrum. A full analysis of all possibilities is beyond the scope of this Letter. 
The most interesting one for the LHC physics would be the two particle peaks plus continuum. 
If the strength of the peak at 125 \gev is not the full standard model one, it is still possible that the Higgs boson satisfies the condition of Eq.~(\ref{mass_vanish}) due to the presence of higher masses in the spectral density.
If we know the strength of the first peak at 125 \gev and the location of the second peak, the strength of the second peak and the form of the continuum are fixed once one extra condition is set.

\begin{figure}
\begin{center}
\includegraphics[width=0.35\textwidth,angle=0]{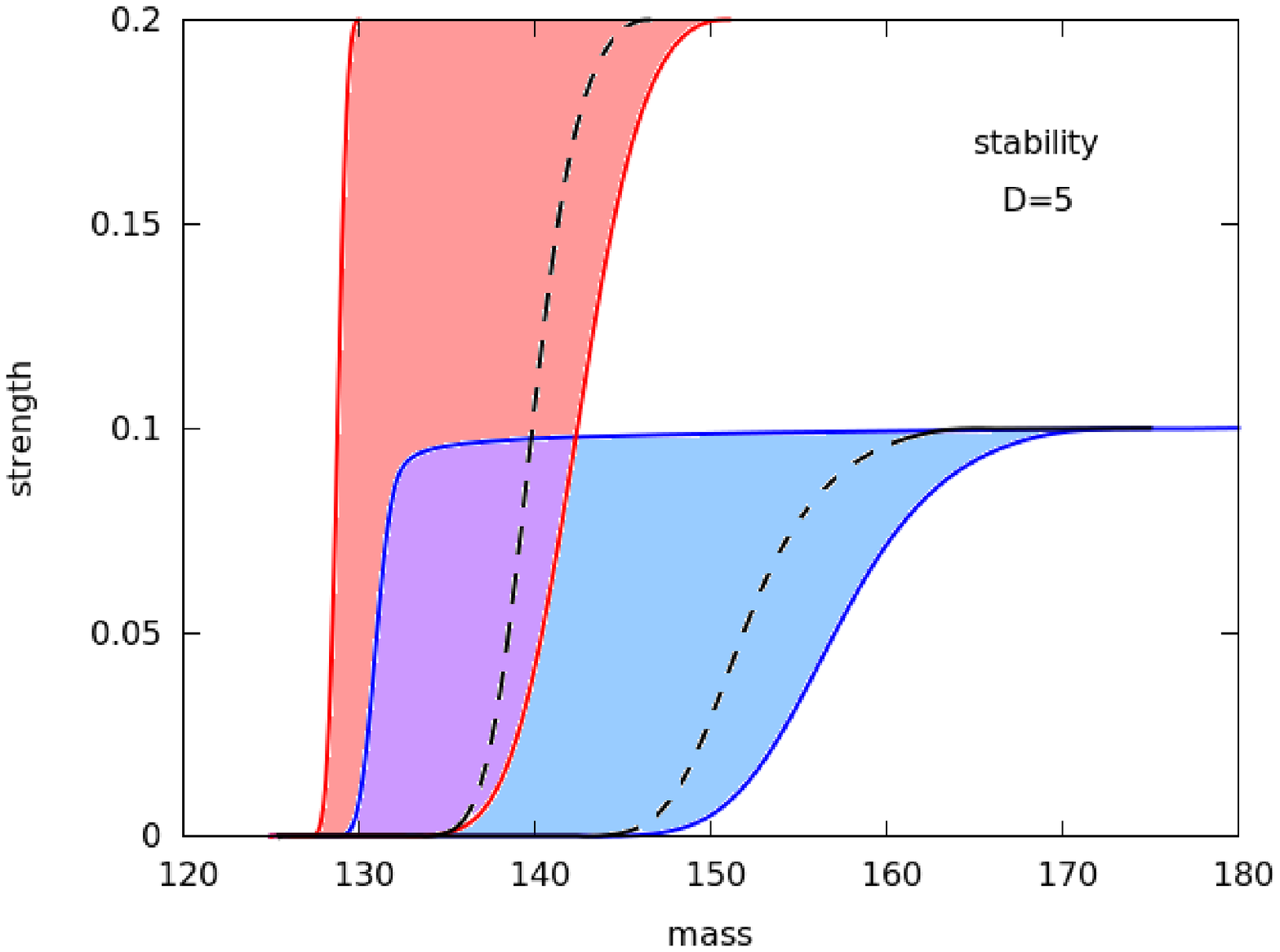}\\
\includegraphics[width=0.35\textwidth,angle=0]{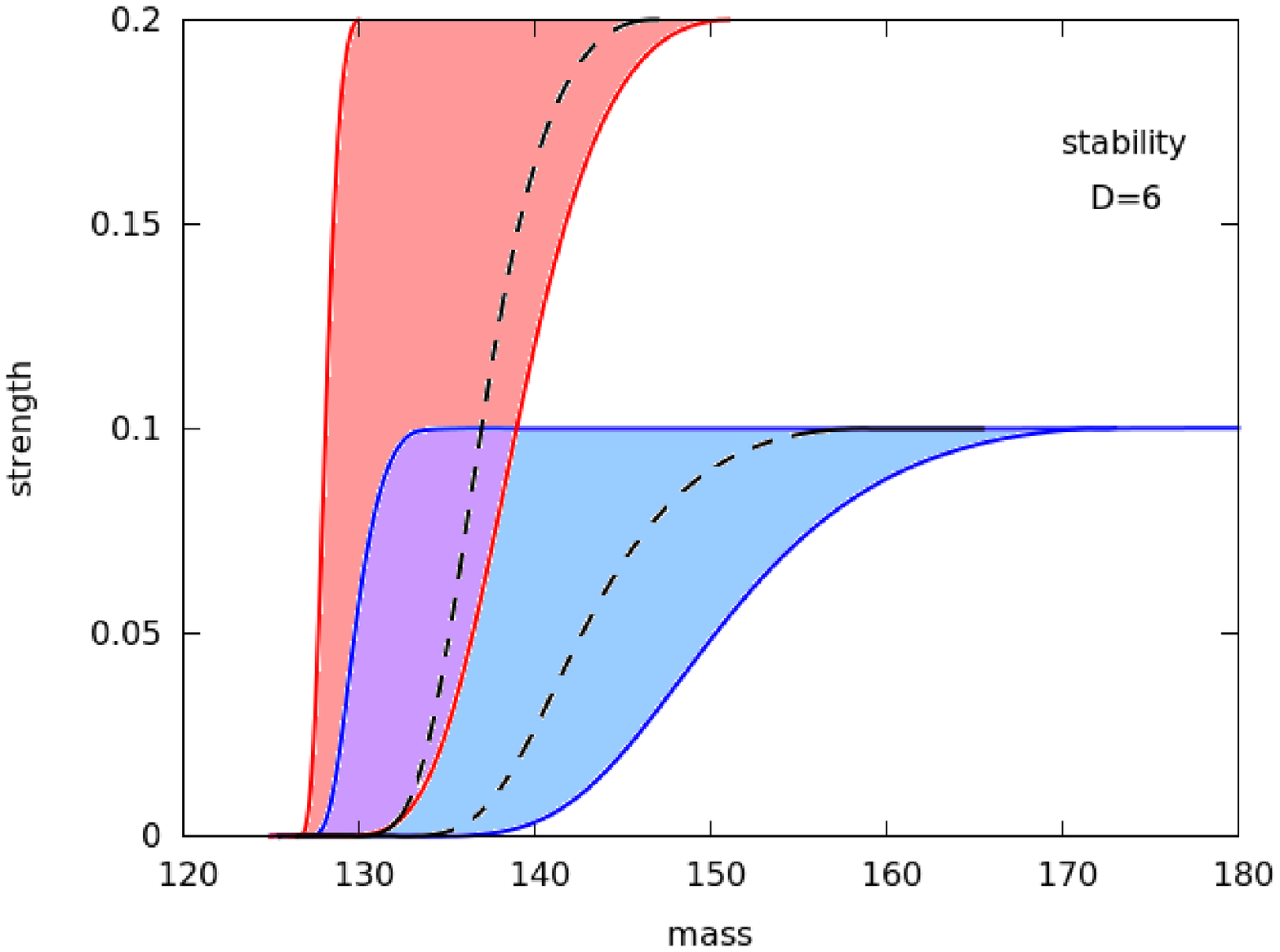}\\
\caption{Strength of the second peak fulfilling the vacuum stability bound for (top) a 5-dimensional and (bottom) a 6-dimensional field, as a function of the mass of the second peak, for an 80\% (left/red) and a 90\% (right/blue) peak at 125 \gev. The dashed lines satisfy Eq.~(\ref{mass_vanish}). The shading is as in Fig.~\ref{fig_dMhMt}.}
\label{fig_contot}
\end{center}
\end{figure}

Here we look at what parameters are allowed by vacuum stability.
We compare it to the case that the  strict condition of Eq.~(\ref{mass_vanish}) is  imposed, i.e., the existence of a conformal window near the Planck scale. 
As an example we give in Fig.~\ref{fig_contot} the strength of the second peak for a 5-dimensional and a 6-dimensional field as a function of the mass of the second peak. We do this for an 80\% and a 90\% peak at 125 \gev.
At the high mass end of the curve one is back at the simple Hill model. We remark that  the area on the {\it right} of the curves is allowed by stability. The curves themselves correspond to the boundary between stability and instability, the curves corresponding to the extremes when varying top and Higgs mass.
The more we increase the suppression of the $125$ GeV Higgs strength (allowing for higher cross sections for the second peak), the more the heavier state needs to be close in mass to the lighter one.
It is clear from the values, that this model is difficult to study at the LHC, especially due to the rather poor experimental resolution. In particular the continuum can probably not be established. 
This type of model can only be studied at a high-luminosity electron-positron machine like the ILC or TLEP, where one can study the Higgs field in the recoil spectrum independent of the decay modes. At least naively a circular collider appears to have advantages for this model, again due to the higher achievable resolution, but the comparison with the ILC should be studied in more detail.

\section{Conclusions}
\label{sect:conclusions}

The one-loop effective potential for the simplest extension of the {\sm} has been used to derive the RGEs for its new parameters. The two-loop {\sm} RGEs have then been completed by the former, to study the stability of the Hill model up to the Planck scale. This is a reasonable approximation, because the evolution of the Higgs quartic coupling reads as in the \sm. Its boundary condition is modified by a positive amount due to the tree-level mixing among the real scalars.

The regions of the parameter space that satisfy the tightest bounds of vacuum stability and triviality were shown, including the impact on the former of varying the Higgs boson and the top quark masses within one sigma around their central value. The resulting area is the one for which the quartic Higgs coupling vanishes at the Planck scale, that some authors in the literature suggest as a minimum requirement for explaining Higgs inflation. 

The experimental constraints on large Higgs boson masses have been discussed, showing that the decay channel into a pair
of  lighter Higgs states should not be neglected. Due to its simplicity, the Hill model can be considered as a benchmark model for the search of any heavy scalar boson, that mixes with the Higgs doublet.
An extension with a higher-dimensional (HEIDI) singlet was discussed, that showed that its physics might be hidden at the
LHC.
In that case a new lepton collider, like the ILC or TLEP, 
is needed to study this model.

\section*{Acknowledgements}
\label{Sec:acknowledgements}

We thank S. Mollet for pointing out an error in an early version of the manuscript.
This work is supported by the Deutsche Forschungsgemeinschaft through the Research Training Group grant
GRK\,1102 \textit{Physics at Hadron Accelerators} and by the Bundesministerium f\"ur Bildung und Forschung within the F\"orderschwerpunkt \textit{Elementary Particle Physics}.


\begin{thebibliography}{}


\bibitem{Higgs:exp} 
  G.~Aad {\it et al.}  [ATLAS Collaboration],
  Phys.\ Lett.\ B {\bf 726}, 88 (2013)
  [arXiv:1307.1427 [hep-ex]],
  [CMS Collaboration],
  CMS-PAS-HIG-13-005.


\bibitem{Bezrukov:2012sa} 
  F.~Bezrukov, M.~Y.~.Kalmykov, B.~A.~Kniehl and M.~Shaposhnikov,
  JHEP {\bf 1210}, 140 (2012)
  [arXiv:1205.2893 [hep-ph]].
  
\bibitem{Degrassi:2012ry}
  G.~Degrassi, S.~Di Vita, J.~Elias-Miro, J.~R.~Espinosa, G.~F.~Giudice, G.~Isidori and A.~Strumia,
  JHEP {\bf 1208} (2012) 098
  [arXiv:1205.6497 [hep-ph]].

\bibitem{Branchina:2013jra} 
  V.~Branchina and E.~Messina,
  Phys.\ Rev.\ Lett.\  {\bf 111} (2013) 241801
  [arXiv:1307.5193 [hep-ph]].


\bibitem{Alekhin:2012py}
  S.~Alekhin, A.~Djouadi and S.~Moch,
  Phys.\ Lett.\ B {\bf 716} (2012) 214
  [arXiv:1207.0980 [hep-ph]].

\bibitem{Masina:2012tz}
  I.~Masina,
  Phys.\ Rev.\ D {\bf 87}, 053001 (2013)
  [arXiv:1209.0393 [hep-ph]].


\bibitem{Extra_scalars}
  T.~Binoth and J.~J.~van der Bij,
  Z.\ Phys.\ C {\bf 75} (1997) 17
  [hep-ph/9608245],
  J.~R.~Espinosa and M.~Quiros,
  Phys.\ Rev.\ D {\bf 76} (2007) 076004
  [hep-ph/0701145],
  S.~Profumo, M.~J.~Ramsey-Musolf and G.~Shaughnessy,
  JHEP {\bf 0708} (2007) 010
  [arXiv:0705.2425 [hep-ph]],
  M.~Bowen, Y.~Cui and J.~D.~Wells,
  JHEP {\bf 0703} (2007) 036
  [hep-ph/0701035].

  \bibitem{Recent_singlets}
  M.~Gonderinger, Y.~Li, H.~Patel and M.~J.~Ramsey-Musolf,
  JHEP {\bf 1001} (2010) 053
  [arXiv:0910.3167 [hep-ph]],
  B.~Batell, S.~Jung and H.~M.~Lee,
  JHEP {\bf 1301} (2013) 135
  [arXiv:1211.2449 [hep-ph]],
  G.~M.~Pruna and T.~Robens,
  Phys.\ Rev.\ D {\bf 88} (2013) 115012
  [arXiv:1303.1150 [hep-ph]].

    
\bibitem{Basso:2013vla} 
  L.~Basso,
  Phys.\ Lett.\ B {\bf 725}, 322 (2013)
  [arXiv:1303.1084 [hep-ph]].
  
\bibitem{Lebedev:2012zw}
  O.~Lebedev,
  Eur.\ Phys.\ J.\ C {\bf 72} (2012) 2058
  [arXiv:1203.0156 [hep-ph]],
  J.~Elias-Miro, J.~R.~Espinosa, G.~F.~Giudice, H.~M.~Lee and A.~Strumia,
  JHEP {\bf 1206} (2012) 031
  [arXiv:1203.0237 [hep-ph]].




\bibitem{vanishing_lambda} 
  F.~Bezrukov and M.~Shaposhnikov,
  JHEP {\bf 0907}, 089 (2009)
  [arXiv:0904.1537 [hep-ph]].
  M.~Shaposhnikov and C.~Wetterich,
  Phys.\ Lett.\ B {\bf 683}, 196 (2010)
  [arXiv:0912.0208 [hep-th]].
  
\bibitem{Hill:1987ea} 
  A.~Hill and J.~J.~van der Bij,
  Phys.\ Rev.\ D {\bf 36}, 3463 (1987).
    

\bibitem{Basso:2012nh}
  L.~Basso, O.~Fischer and J.~J.~van der Bij,
  Europhys.\ Lett.\ {\bf 101} (2013) 51004
  [arXiv:1212.5560 [hep-ph]].


\bibitem{Chatrchyan:2013yoa} 
  S.~Chatrchyan {\it et al.}  [CMS Collaboration],
  Eur.\ Phys.\ J.\ C {\bf 73}, 2469 (2013)
  [arXiv:1304.0213 [hep-ex]].

\bibitem{invisible}
  G.~Belanger, B.~Dumont, U.~Ellwanger, J.~F.~Gunion and S.~Kraml,
  Phys.\ Lett.\ B {\bf 723}, 340 (2013)
  [arXiv:1302.5694 [hep-ph]],
  P.~P.~Giardino, K.~Kannike, I.~Masina, M.~Raidal and A.~Strumia,
  arXiv:1303.3570 [hep-ph],
  A.~Falkowski, F.~Riva and A.~Urbano,
  JHEP {\bf 1311}, 111 (2013)
  [arXiv:1303.1812 [hep-ph]],
  J.~Ellis and T.~You,
  JHEP {\bf 1306}, 103 (2013)
  [arXiv:1303.3879 [hep-ph]].

\bibitem{Djouadi:1997yw}
  A.~Djouadi, J.~Kalinowski and M.~Spira,
  Comput.\ Phys.\ Commun.\  {\bf 108} (1998) 56
  [hep-ph/9704448].
    
\bibitem{vanderBij:2006ne} 
  J.~J.~van der Bij,
  Phys.\ Lett.\ B {\bf 636}, 56 (2006)
  [hep-ph/0603082].

\bibitem{vanderBij:2011wy}
  J.~J.~van der Bij and B.~Pulice,
  Nucl.\ Phys.\ B {\bf 853} (2011) 49
  [arXiv:1104.2062 [hep-ph]].



\end{thebibliography}
\end{document}